\documentclass[10pt,twocolumn]{article}

\usepackage[utf8]{inputenc}
\usepackage[T1]{fontenc}
\usepackage{lmodern}
\usepackage{microtype}
\usepackage[top=1in,bottom=1in,left=0.85in,right=0.85in,columnsep=0.28in]{geometry}
\usepackage{amsmath,amssymb}
\usepackage{booktabs}
\usepackage{multirow}
\usepackage{xcolor}
\usepackage[colorlinks=true,linkcolor=blue!60!black,citecolor=blue!60!black,urlcolor=blue!60!black]{hyperref}
\usepackage{pgfplots}
\usepackage{pgfplotstable}
\pgfplotsset{compat=1.18}
\usepackage{caption}
\usepackage{subcaption}
\usepackage{cleveref}
\usepackage{enumitem}
\usepackage{url}
\usepackage{xurl}          
\usepackage{titlesec}
\usepackage{abstract}
\definecolor{myblue}{RGB}{31,119,180}
\definecolor{myorange}{RGB}{255,127,14}
\definecolor{myred}{RGB}{200,40,40}
\definecolor{mygray}{RGB}{180,180,180}
\definecolor{mygreen}{RGB}{44,160,44}
\definecolor{mypurple}{RGB}{148,103,189}

\titleformat{\section}{\normalfont\normalsize\bfseries\scshape}{\thesection.}{0.5em}{}
\titleformat{\subsection}{\normalfont\normalsize\bfseries}{\thesubsection.}{0.4em}{}
\titleformat{\subsubsection}{\normalfont\normalsize\itshape}{\thesubsubsection.}{0.4em}{}
\titlespacing*{\section}{0pt}{6pt plus 2pt minus 1pt}{3pt plus 1pt}
\titlespacing*{\subsection}{0pt}{4pt plus 1pt}{2pt plus 1pt}

\title{\vspace{-1.5em}\textbf{Polars inside Intel SGX2 Enclaves:\
An Empirical Study of Confidential Analytical Query Processing}}
\author{%
  Wei Wang\\
  \textit{Anonym, Mozilla}\\
  \href{mailto:wewang@mozilla.com}{\texttt{wewang@mozilla.com}}
  \and
  Burns Smith\\
  \textit{Anonym, Mozilla}\\
  \href{mailto:busmith@mozilla.com}{\texttt{busmith@mozilla.com}}
  \and
  Kenny Leftin\\
  \textit{Anonym, Mozilla}\\
  \href{mailto:kleftin@mozilla.com}{\texttt{kleftin@mozilla.com}}
}
\date{}

\begin{document}

\maketitle
\thispagestyle{empty}

\begin{abstract}
\noindent
Trusted Execution Environments (TEEs) have renewed interest in confidential analytics, but most prior evaluations focus on SQL database engines or earlier SGX generations. This paper studies an Arrow-native DataFrame engine, Polars, running inside Intel SGX2 enclaves via Gramine on TPC-H SF30 with Azure Blob Storage. We report both the standard TPC-H power score and a query-only variant that removes table-loading time in order to separate compute overhead from data-ingestion overhead. Across four dataset-width configurations (approximately 22--73\,GB), end-to-end overhead remains nearly constant at 1.49--1.56$\times$, but this composite metric obscures two distinct behaviors: query-only overhead declines from 1.51--1.52$\times$ to 1.43--1.44$\times$, whereas table-loading overhead rises from 2.27$\times$ to 4.07$\times$. We further show that overhead is not uniform across queries: for the len130 configuration, the median per-query SGX slowdown is 1.45$\times$ with a maximum of 2.57$\times$, and a small set of queries exhibits pronounced run-to-run spikes consistent with stateful EPC pressure. Finally, we compare Polars' lazy and eager APIs under the same TEE setting. Lazy execution is 2.25--2.27$\times$ faster overall, while eager execution fails with out-of-memory errors at 41\,GB and above. Relative to the recent DuckDB-SGX2 study, our results suggest that SGX2 can support Arrow-native analytical processing with a similar order of security overhead, but that load-path amplification and API-level optimization are first-order determinants of end-to-end performance.
\end{abstract}

\section{Introduction}

Confidential computing aims to protect \emph{data in use} by executing applications inside hardware-isolated trusted execution environments (TEEs). Among commodity CPU TEEs, Intel Software Guard Extensions (SGX) exposes a process-level enclave abstraction in which code and data are protected against a privileged but untrusted system software stack~\cite{costan2016sgx,sgx2dmm}. While early SGX deployments were severely constrained by enclave memory capacity, second-generation SGX hardware (SGX2) adds dynamic memory management and substantially larger protected memory budgets, renewing interest in confidential analytical processing~\cite{sgx2dmm,elhindi2022sgx2}.

The database community has already shown that SGX can support secure query processing, but the literature is dominated by SQL systems or SGX1-era prototypes such as Opaque, EnclaveDB, StealthDB, and EncDBDB~\cite{zheng2017opaque,priebe2018enclavedb,gribov2019stealthdb,fuhry2021encdbdb}. The closest prior work to ours is DuckDB-SGX2, which evaluates an embedded analytical DBMS on TPC-H SF30 and shows that a carefully tuned SGX2 deployment can keep overhead below 2$\times$ while remaining sensitive to paging, NUMA locality, and allocator behavior~\cite{battiston2024duckdb}. However, another increasingly important analytical execution model \emph{DataFrame-first}: systems such as Polars and DataFusion expose Arrow-native operators, lazy optimization, and embeddable runtimes that are widely used in ETL, feature engineering, and analytical application backends~\cite{polarsdocs,polarslazy,polarsopt,lamb2024datafusion,arrowdocs}.

This paper studies that design point directly. We evaluate Polars running inside SGX2 via Gramine on Azure DCsv3 hardware using TPC-H SF30 data stored in Azure Blob Storage. The goal is not merely to report a single slowdown number, but to identify which parts of the execution path degrade under SGX2, how the degradation changes with dataset width, how stable the behavior is across repeated runs, and whether Polars' lazy and eager APIs interact differently with TEE overhead.

Our study makes four contributions. First, it provides a detailed SGX2 characterization of an Arrow-native DataFrame engine rather than a SQL engine. Second, it decomposes end-to-end overhead into query execution and data-loading components, showing that a stable headline overhead hides divergent scaling trends. Third, it reports per-query and run-level behavior, revealing that most queries are stable while a small number experience severe outliers consistent with EPC pressure. Fourth, it compares lazy and eager execution under the same enclave setting and shows that execution mode has larger practical impact than the SGX tax itself.

\section{Background and Related Work}

\subsection{Intel SGX2 and Gramine}

Intel SGX protects an application's enclave pages inside the Enclave Page Cache (EPC), with access checks and memory encryption enforced by hardware~\cite{costan2016sgx}. SGX2 extends SGX with dynamic memory management, enabling enclaves to grow after creation rather than fixing all memory up front~\cite{sgx2dmm}. Empirical studies on second-generation hardware show that SGX2 alleviates some SGX1 bottlenecks but does not eliminate high penalties from cache misses and EPC paging~\cite{elhindi2022sgx2}. 

We rely on Gramine, a library OS that allows largely unmodified Linux applications to run inside SGX enclaves~\cite{tsai2017graphene,gramine}. This choice is important for analytical software stacks because it preserves existing runtimes, language bindings, and I/O libraries, but it also means that enclave performance still reflects interaction with the operating system, the allocator, and the storage stack.

\subsection{Polars, Arrow, and Lazy Optimization}

Polars is a Rust-based DataFrame library built around columnar execution and Apache Arrow-compatible memory layouts~\cite{polarsdocs,arrowdocs}. It exposes two principal execution modes. In the \emph{eager} API, operators execute immediately. In the \emph{lazy} API, operators are compiled into a plan that can be optimized before execution. Polars documents predicate pushdown, projection pushdown, slice pushdown, and common-subplan elimination as core optimizations in lazy mode~\cite{polarslazy,polarsopt}. These optimizations are particularly relevant in TEEs because they can reduce both the total bytes read into enclave-protected memory and the peak size of intermediate results.

The broader trend toward Arrow-native embedded analytics is also visible in systems such as DataFusion, which positions Arrow as the execution and interchange substrate for modular analytical engines~\cite{lamb2024datafusion}. Polars differs from SQL engines in API surface and optimizer interface, so its SGX2 behavior cannot be inferred directly from prior DuckDB-based results.

\subsection{Confidential Data Management}

Prior secure data-management systems have explored several points in the design space. Opaque extends Spark with secure and oblivious operators for distributed analytics~\cite{zheng2017opaque}. EnclaveDB secures database execution using SGX while targeting confidentiality, integrity, and freshness~\cite{priebe2018enclavedb}. StealthDB focuses on encrypted transactional processing with limited DBMS changes~\cite{gribov2019stealthdb}. EncDBDB targets read-oriented analytical processing in a column-oriented, compressed in-memory design using enclaves~\cite{fuhry2021encdbdb}. These systems demonstrate the feasibility of secure query processing, but they were largely designed around SGX1-era constraints or different execution abstractions.

\subsection{Relation to DuckDB-SGX2}

DuckDB-SGX2 is the closest reference point for our study~\cite{battiston2024duckdb}. Battiston et al. report that a well-configured DuckDB deployment on TPC-H SF30 incurs roughly 1.5--2$\times$ overhead, and they identify several hazards that dominate performance when configuration is poor, including EPC paging, loss of NUMA locality, and allocator-induced fragmentation. Their study also combines SGX2 execution with Parquet modular encryption for data at rest~\cite{battiston2024duckdb,parquet_encryption}. 

\section{System and Methodology}

\subsection{Implementation}

We run Polars inside an SGX2 enclave using Gramine~\cite{tsai2017graphene,gramine}. The application process is standard Python plus Polars, with system interfaces mediated by the library OS. Input data resides in Azure Blob Storage and is fetched over HTTPS before being decoded into in-memory columnar structures. The experimental platform is Azure's \texttt{Standard\_DC48s\_v3} VM, which provides 48 vCPUs, 384\,GiB DRAM, and 256\,GiB encrypted enclave memory capacity (EPC) on SGX-capable hardware~\cite{azuredcsv3}. The non-SGX baseline runs on the same VM family outside enclave mode.

\subsection{Workload and Dataset Variants}

We use TPC-H SF30, which consists of 22 decision-support queries over eight tables~\cite{tpch}. The dataset contains approximately 180 million \texttt{lineitem} rows. To study sensitivity to input width, we generate four variants by increasing the length of all \texttt{*\_comment} columns, producing approximate dataset sizes of 22, 30, 41, and 73\,GB (\Cref{tab:configs}).

\begin{table}[h]
\centering
\caption{Dataset configurations.}
\label{tab:configs}
\small
\begin{tabular}{lrr}
\toprule
Config & \texttt{comment\_len} & Approx.\ size \\
\midrule
len80  &  80 chars & $\approx$22\,GB \\
len130 & 130 chars & $\approx$30\,GB \\
len200 & 200 chars & $\approx$41\,GB \\
len400 & 400 chars & $\approx$73\,GB \\
\bottomrule
\end{tabular}
\end{table}

This manipulation should be interpreted carefully. It keeps row counts fixed and primarily increases scan, transport, and decode volume, but it is \emph{not} a pure I/O knob because TPC-H does include predicates over comment fields in some queries (notably Q13 and Q16)~\cite{boncz2013tpch,boncz2013tpch}. We therefore treat these variants as controlled workload-width changes rather than claiming they are completely orthogonal to query semantics.

\subsection{Measurement Protocol and Metrics}

Each configuration is executed five times per query, with the first run treated as warm-up and excluded from aggregate statistics. We report the TPC-H power score
\[
\mathit{PS} = 3600 \times \mathit{SF}\,/\,\hat{G}(t_i),
\]
where $\hat{G}$ is the geometric mean over the component times and higher is better~\cite{tpch}. Because the standard power score includes both the 22 query timings and table-loading time, we additionally report a \emph{query-only} power score computed over the query timings alone. This allows us to distinguish query-execution overhead from data-loading overhead.

For throughput metrics, we define SGX overhead as
\[
R_{PS} = \mathit{PS}_{\text{non-SGX}} / \mathit{PS}_{\text{SGX}}.
\]
For latency metrics such as load time and per-query time, we define overhead as
\[
R_t = t_{\text{SGX}} / t_{\text{non-SGX}}.
\]
We report both mean-based aggregate metrics and distributional summaries when discussing individual queries.

\section{Results}

\subsection{End-to-End Throughput and Overhead Decomposition}

\Cref{tab:power} reports both the standard TPC-H power score and the query-only variant. The main observation is that a nearly constant end-to-end overhead does not imply a constant overhead source. Across the four configurations, end-to-end overhead stays within a narrow 1.49--1.56$\times$ range, but query-only overhead drops from 1.51--1.52$\times$ for the smaller datasets to 1.43--1.44$\times$ for the larger ones.

\begin{table}[h]
\centering
\caption{TPC-H power scores (lazy API). The \emph{masking gap} is the difference between end-to-end and query-only overhead.}
\label{tab:power}
\small
\setlength{\tabcolsep}{3pt}
\begin{tabular}{lrrrrr}
\toprule
Dataset & Non-SGX & SGX & E2E OH & Query OH & Gap \\
\midrule
len80  & 176{,}542 & 115{,}167 & 1.53$\times$ & 1.51$\times$ & 0.02 \\
len130 & 173{,}759 & 111{,}640 & 1.56$\times$ & 1.52$\times$ & 0.04 \\
len200 & 171{,}995 & 115{,}605 & 1.49$\times$ & 1.43$\times$ & 0.06 \\
len400 & 166{,}360 & 110{,}438 & 1.51$\times$ & 1.44$\times$ & 0.07 \\
\bottomrule
\end{tabular}
\end{table}

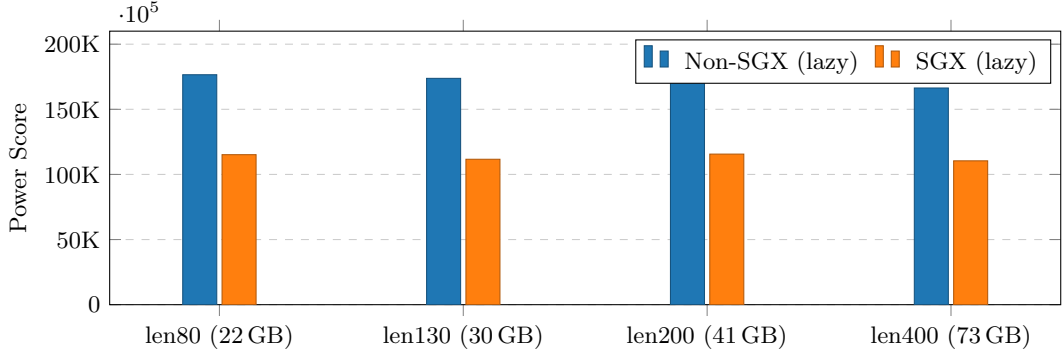
\begin{figure*}[t]
\centering
\begin{tikzpicture}
\begin{axis}[
  ybar,
  bar width=0.45cm,
  width=0.82\textwidth,
  height=5.2cm,
  enlarge x limits=0.15,
  legend style={at={(0.99,0.97)}, anchor=north east, font=\small,
                legend columns=2, column sep=0.4em},
  ylabel={Power Score},
  ylabel style={font=\small},
  xtick={1,2,3,4},
  xticklabels={len80 (22\,GB), len130 (30\,GB), len200 (41\,GB), len400 (73\,GB)},
  xticklabel style={font=\small},
  yticklabel style={font=\small},
  ymin=0, ymax=210000,
  ytick={0,50000,100000,150000,200000},
  yticklabels={0,50K,100K,150K,200K},
  ymajorgrids=true,
  grid style={dashed,gray!40},
  tick label style={font=\small},
]
\addplot[fill=myblue,draw=myblue!70!black] coordinates {
  (1,176542)(2,173759)(3,171995)(4,166360)
};
\addplot[fill=myorange,draw=myorange!70!black] coordinates {
  (1,115167)(2,111640)(3,115605)(4,110438)
};
\legend{Non-SGX (lazy), SGX (lazy)}
\end{axis}
\end{tikzpicture}
\caption{TPC-H power scores for the Polars lazy API across four dataset widths. The aggregate overhead is stable, but \Cref{tab:power} shows that this stability masks diverging query and load-path behavior.}
\label{fig:power}
\end{figure*}

\Cref{fig:power} visualizes the power scores. The non-SGX score decreases by only 5.8\% from len80 to len400, while the SGX score decreases by 4.1\%. At the level of the composite metric, the system therefore appears remarkably stable. The widening masking gap in \Cref{tab:power} shows why that interpretation is incomplete: larger datasets make the loading path relatively worse while leaving the query engine itself slightly better amortized.

\subsection{Load-Path Amplification}

\Cref{tab:loadtimes} isolates table-loading time. In contrast to the relatively stable query-only overhead, loading overhead increases monotonically from 2.27$\times$ to 4.07$\times$. Non-SGX load time scales by about 2.08$\times$ from len80 to len400, whereas SGX load time scales by about 3.73$\times$ over the same range.

\begin{table}[h]
\centering
\caption{Table load times and SGX overhead.}
\label{tab:loadtimes}
\small
\begin{tabular}{lrrr}
\toprule
Dataset & Non-SGX (s) & SGX (s) & Overhead \\
\midrule
len80  &  92.5 & 209.9 & 2.27$\times$ \\
len130 & 123.3 & 330.1 & 2.68$\times$ \\
len200 & 134.1 & 446.0 & 3.33$\times$ \\
len400 & 192.5 & 783.7 & 4.07$\times$ \\
\bottomrule
\end{tabular}
\end{table}

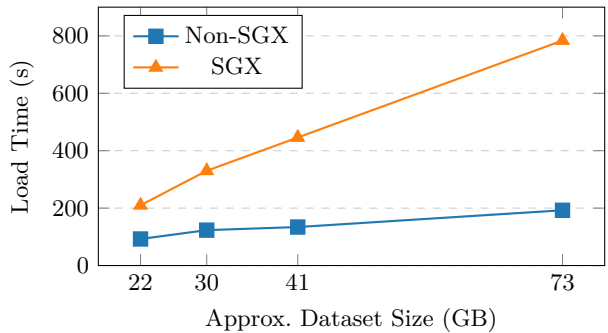
\begin{figure}[h]
\centering
\begin{tikzpicture}
\begin{axis}[
  width=\columnwidth,
  height=5.0cm,
  xlabel={Approx.\ Dataset Size (GB)},
  ylabel={Load Time (s)},
  xlabel style={font=\small},
  ylabel style={font=\small},
  xticklabel style={font=\small},
  yticklabel style={font=\small},
  xtick={22,30,41,73},
  xticklabels={22,30,41,73},
  ymin=0, ymax=900,
  ymajorgrids=true,
  grid style={dashed,gray!40},
  legend style={at={(0.05,0.97)}, anchor=north west, font=\small},
  mark size=2.5pt,
]
\addplot[color=myblue, mark=square*, thick] coordinates {
  (22,92.5)(30,123.3)(41,134.1)(73,192.5)
};
\addplot[color=myorange, mark=triangle*, thick] coordinates {
  (22,209.9)(30,330.1)(41,446.0)(73,783.7)
};
\legend{Non-SGX, SGX}
\end{axis}
\end{tikzpicture}
\caption{Table-loading time versus dataset size. The SGX path scales super-linearly relative to the baseline.}
\label{fig:loadtime}
\end{figure}

The most plausible explanation is that loading stresses the EPC much more severely than the steady-state query plans do. During load, Polars must materialize decoded Arrow buffers while continuously moving bytes from remote object storage through the protected address space. When the active footprint exceeds available EPC, the system pays the cost of enclave paging and page revalidation rather than ordinary memory traffic~\cite{costan2016sgx,elhindi2022sgx2}. This interpretation is also consistent with DuckDB-SGX2, which identifies EPC paging as the dominant source of configuration-sensitive slowdowns~\cite{battiston2024duckdb}.

The practical consequence is that data loading dominates end-to-end behavior. For the len130 configuration, the aggregate query-compute time is 24.7\,s outside SGX and 38.1\,s inside SGX, whereas table loading alone takes 123.3\,s and 330.1\,s, respectively. Thus, the load stage accounts for roughly 83\% of end-to-end time outside SGX and 90\% inside SGX for this configuration.

\subsection{Per-Query Overhead Distribution}

\begin{figure*}[t]
\centering
\begin{tikzpicture}
\begin{axis}[
  ybar,
  bar width=0.36cm,
  width=0.88\textwidth,
  height=5.2cm,
  xlabel={TPC-H Query},
  ylabel={SGX Overhead ($\times$)},
  xlabel style={font=\small},
  ylabel style={font=\small},
  xtick={1,...,22},
  xticklabels={Q1,Q2,Q3,Q4,Q5,Q6,Q7,Q8,Q9,Q10,Q11,Q12,Q13,Q14,Q15,Q16,Q17,Q18,Q19,Q20,Q21,Q22},
  xticklabel style={rotate=45, anchor=east, font=\scriptsize},
  yticklabel style={font=\small},
  ymin=1.0, ymax=3.0,
  ytick={1.0,1.25,1.50,1.75,2.00,2.25,2.50,2.75,3.00},
  ymajorgrids=true,
  grid style={dashed,gray!40},
  enlarge x limits=0.03,
]
\addplot[fill=myblue!80, draw=myblue!60!black] coordinates {
  (1,1.862)(2,1.833)(3,1.502)(4,2.569)(5,1.363)(6,1.566)
  (7,1.658)(8,1.445)(9,1.244)(10,1.401)(11,1.290)(12,1.363)
  (13,1.619)(14,1.567)(15,1.428)(16,1.484)(17,1.456)(18,1.450)
  (19,1.455)(20,1.442)(21,1.387)(22,1.417)
};
\draw[dashed, myred, thick, line width=1pt] (axis cs:0,1.52) -- (axis cs:23,1.52);
\node[myred, font=\scriptsize, anchor=west] at (axis cs:22.3,1.57) {geo. mean 1.52$\times$};
\end{axis}
\end{tikzpicture}
\caption{Per-query SGX overhead for len130. Most queries cluster around moderate slowdown, while Q04 is a clear outlier.}
\label{fig:perquery}
\end{figure*}
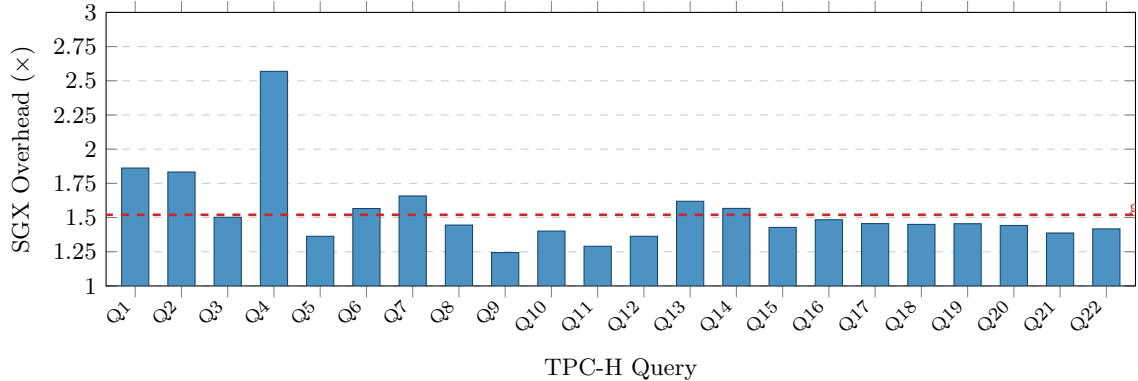

Aggregate power scores hide meaningful per-query variation. \Cref{fig:perquery} reports the len130 per-query SGX overheads. The geometric-mean overhead for this configuration is 1.52$\times$, but the arithmetic median across the 22 queries is 1.45$\times$ and the interquartile range is approximately 1.41--1.57$\times$. Eighteen of the 22 queries fall in the 1.25--1.66$\times$ band, indicating that most of the workload clusters around a moderate slowdown rather than the worst-case outlier.

Three aspects of the distribution are noteworthy. First, Q04 is a clear outlier at 2.57$\times$. Second, Q01 and Q02 are elevated at 1.86$\times$ and 1.83$\times$, respectively, indicating that some scan-heavy or aggregation-heavy operators remain noticeably more sensitive to enclave execution than the median query. Third, the minimum observed overhead is still above 1.0$\times$ (Q09 at 1.24$\times$), suggesting that the workload never enters a regime where SGX costs are completely hidden by other bottlenecks.

\subsection{Run-to-Run Instability and Stateful EPC Pressure}

The average slowdown of Q04 is not solely a matter of constant overhead; it is driven by pronounced run-to-run instability. For the len130 SGX configuration, most queries are stable, but a small set exhibits large coefficients of variation: Q04 (47\%), Q07 (47\%), and Q02 (48\%). \Cref{tab:spikes} shows the per-run times for Q04 and Q07.

\begin{table}[h]
\centering
\caption{Run-by-run SGX latency (seconds) for len130. Run 3 produces a simultaneous spike in both queries.}
\label{tab:spikes}
\small
\setlength{\tabcolsep}{4pt}
\begin{tabular}{lrrrrr}
\toprule
 & Run 1 & Run 2 & Run 3 & Run 4 & Run 5 \\
\midrule
Q04 & 2.81 & 2.66 & 6.09 & 2.48 & 2.56 \\
Q07 & 2.88 & 2.85 & 6.76 & 2.88 & 2.93 \\
\bottomrule
\end{tabular}
\end{table}

The fact that Q04 and Q07 spike together in the same run argues against a purely query-local explanation. Similar behavior is also reported elsewhere in the dataset family: for example, a large Q07 spike appears in len80 run 2, and a Q04 spike appears in len400 run 1. The spike position is therefore not tied to a fixed query index. A more plausible interpretation is that EPC pressure accumulates across the benchmark sequence within the same long-lived process and occasionally triggers a paging cascade when a subsequent query demands additional memory. DuckDB-SGX2 reports a related phenomenon in which repeated execution of hash-heavy queries deteriorates after allocator and memory-state interactions build up over time~\cite{battiston2024duckdb}. 

This result matters methodologically as well as operationally. If benchmarking is performed with one process per query, these spikes may be underrepresented; if long-lived analytical services execute many queries in the same runtime, they may be more common than median-query numbers suggest.

\subsection{Lazy versus Eager Execution}

We next compare Polars' lazy and eager APIs. \Cref{tab:eagerpower} shows the overall power score at len80, the only configuration for which the eager SGX run completes successfully. Lazy execution outperforms eager execution by 2.27$\times$ outside SGX and 2.25$\times$ inside SGX. The corresponding SGX overhead for eager execution, 1.52$\times$, is almost identical to the lazy-api overhead at len80, indicating that SGX does not erase the optimizer's benefits; instead, the optimization and the SGX tax compound.

\begin{table}[h]
\centering
\caption{Power scores for lazy and eager execution. Eager SGX runs fail with out-of-memory errors at len200 and above.}
\label{tab:eagerpower}
\small
\begin{tabular}{lrrrr}
\toprule
 & \multicolumn{2}{c}{Non-SGX} & \multicolumn{2}{c}{SGX} \\
\cmidrule(lr){2-3}\cmidrule(lr){4-5}
Dataset & Lazy & Eager & Lazy & Eager \\
\midrule
len80  & 176{,}542 & 77{,}887 & 115{,}167 & 51{,}136 \\
\bottomrule
\end{tabular}
\end{table}

The load path also penalizes eager execution. At len80, eager loading is slightly faster than lazy loading outside SGX (88.4\,s versus 92.5\,s), but inside SGX it becomes slower (237\,s versus 209.9\,s). This reversal is consistent with the absence of lazy-mode pruning during scan and materialization.

\begin{figure*}[t]
\centering
\begin{tikzpicture}
\begin{axis}[
  ybar,
  bar width=0.36cm,
  width=0.88\textwidth,
  height=5.4cm,
  xlabel={TPC-H Query},
  ylabel={Eager / Lazy slowdown ($\log_2$ scale)},
  xlabel style={font=\small},
  ylabel style={font=\small},
  xtick={1,...,22},
  xticklabels={Q1,Q2,Q3,Q4,Q5,Q6,Q7,Q8,Q9,Q10,Q11,Q12,Q13,Q14,Q15,Q16,Q17,Q18,Q19,Q20,Q21,Q22},
  xticklabel style={rotate=45, anchor=east, font=\scriptsize},
  yticklabel style={font=\small},
  ymode=log,
  log basis y=2,
  ymin=0.6, ymax=200,
  ymajorgrids=true,
  grid style={dashed,gray!40},
  enlarge x limits=0.03,
  ytick={0.5,1,2,4,8,16,32,64,128},
  yticklabels={$0.5\times$,$1\times$,$2\times$,$4\times$,$8\times$,$16\times$,$32\times$,$64\times$,$128\times$},
]
\addplot[fill=mygreen!70, draw=mygreen!60!black] coordinates { (1,0.765) };
\addplot[fill=myblue!70, draw=myblue!60!black] coordinates {
  (2,3.149)(3,3.344)(4,1.267)(5,2.038)(6,2.172)
  (7,4.417)(8,3.182)
};
\addplot[fill=myred!90, draw=myred!80!black] coordinates { (9,5.175) };
\addplot[fill=myblue!70, draw=myblue!60!black] coordinates {
  (10,2.139)(11,2.844)(12,1.268)(13,1.019)(14,2.096)
  (15,1.109)(16,3.556)(17,1.657)(18,1.182)
};
\addplot[fill=myred!90, draw=myred!80!black] coordinates { (19,121.693) };
\addplot[fill=myblue!70, draw=myblue!60!black] coordinates {
  (20,1.129)(21,2.624)(22,1.028)
};
\draw[dashed, black!60, line width=0.8pt] (axis cs:0,1) -- (axis cs:23,1);
\node[font=\scriptsize\bfseries, myred, anchor=south] at (axis cs:19,140) {121$\times$};
\end{axis}
\end{tikzpicture}
\caption{Per-query eager/lazy slowdown for len80 on a $\log_2$ scale. The optimizer effect is highly non-uniform and dominated by a small number of queries.}
\label{fig:eager}
\end{figure*}
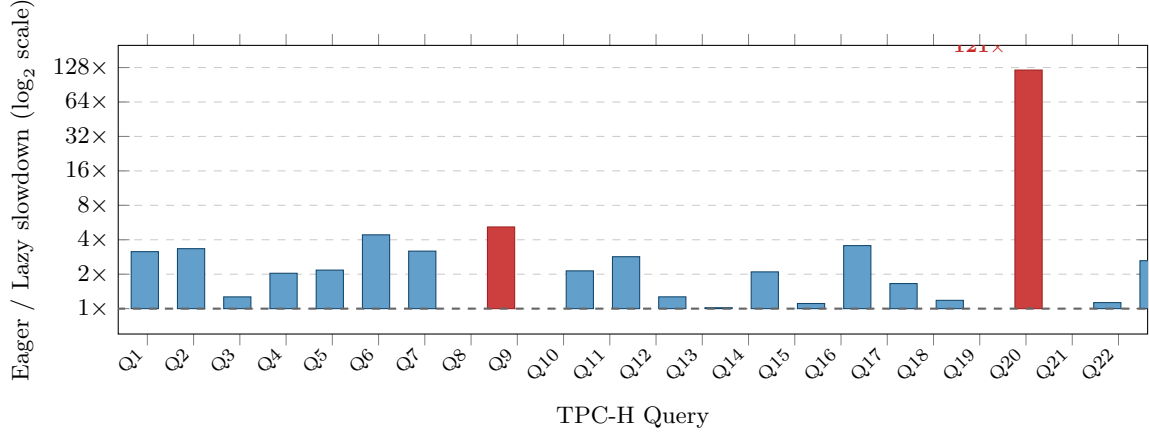

\Cref{fig:eager} reports the per-query eager/lazy slowdown at len80. Most queries suffer moderate regression, but the effect is highly skewed. Q19 is the dominant outlier at 121$\times$, and Q09 also shows a large 5.2$\times$ slowdown. At the other end of the distribution, Q01 is slightly faster in eager mode (0.77$\times$), suggesting that some simple scan-aggregate patterns derive little advantage from lazy planning in this workload.

The main takeaway is not merely that lazy is faster, but that lazy execution is substantially more robust under enclave memory pressure. Eager execution fails at 41\,GB and above, whereas the lazy path completes all four configurations. For confidential analytics, optimizer effectiveness is therefore also a memory-management mechanism.

\section{Threats to Validity}

Our experiments use a single hardware and software stack: Azure DCsv3 hardware, Gramine, Azure Blob Storage, and one Polars-based implementation. Results may differ on other SGX2 platforms, other enclave runtimes, or local-storage configurations. The measurements also focus on the power-style single-stream benchmark rather than concurrent throughput.

The dataset-width manipulation is intentionally controlled but imperfect. Because some TPC-H queries reference comment fields, widening those fields is not purely a scan-only perturbation~\cite{boncz2013tpch,boncz2013tpch}. This does not invalidate the experiment, but it means the results should be read as workload-variant sensitivity rather than as an isolated I/O microbenchmark.

Finally, our analysis identifies likely mechanisms such as EPC paging and stateful memory pressure, but it does not instrument the enclave runtime with page-fault or allocator telemetry. The claims are therefore grounded in the timing evidence and prior SGX literature rather than direct page-level traces.

\section{Conclusion}

This paper evaluated Polars inside Intel SGX2 enclaves on TPC-H SF30. The main result is that SGX2 overhead for Arrow-native analytical processing is moderate but structurally heterogeneous. The standard end-to-end power score suggests a nearly constant 1.49--1.56$\times$ overhead, yet a deeper decomposition shows that query execution itself improves slightly with dataset width while the loading path becomes progressively more expensive. Most queries cluster around a moderate slowdown, but a small number of outliers display substantial variance, and eager execution magnifies both runtime and memory pressure.


\end{document}